\documentclass[reprint,amsmath,amssymb,aps,pra]{revtex4-1}
\usepackage{graphicx}% Include figure files
\usepackage{dcolumn}% Align table columns on decimal point
\usepackage{mathrsfs,amsmath}
\usepackage{bm}
\usepackage{xcolor}
\newcommand{\beq}{\begin{equation}}
\newcommand{\eneq}{\end{equation}}
\renewcommand{\vec}[1]{\mathbf{#1}}

\begin{document}

\title{Uniaxial modulation and the Berezinskii-Kosterlitz-Thouless transition}

\author{Domenico Giuliano$^{(1,2)}$, Phong H. Nguyen$^{(3,4)}$, Andrea Nava$^{(1,2,5)}$, and Massimo Boninsegni$^{(3)}$ }
\affiliation{
$^{(1)}$Dipartimento di Fisica, Universit\`a della Calabria Arcavacata di 
Rende I-87036, Cosenza, Italy \\
$^{(2)}$I.N.F.N., Gruppo collegato di Cosenza, 
Arcavacata di Rende I-87036, Cosenza, Italy \\
$^{(3)}$Department of Physics, University of Alberta, Edmonton, T6G 2E1, Alberta, Canada \\
$^{(4)}$ Faculty of Physics, VNU University of Science, Vietnam National University, 334 Nguyen Trai Street, Thanh Xuan, Hanoi, Vietnam\\
$^{(5)}$ Institut f\"ur Theoretische Physik, Heinrich-Heine-Universit\"at, D-40225 D\"usseldorf, Germany}
\date{today}

\begin{abstract}
We present a theoretical study of the Berezinskii-Kosterlitz-Thouless transition of a two-dimensional superfluid in the presence of an externally imposed density modulation along a single axis. The subject is investigated in the context of the $|\psi|^4$ classical field theory, by means of analytical and numerical techniques. We show that, as the amplitude of the modulation increases, the physics of the system approaches that of the anisotropic $x$-$y$ model, with a suppressed superfluid transition temperature and an anisotropic response, but with no dimensional crossover.
\end{abstract}
\maketitle

\section{Introduction}
\label{intro}
The intriguing behavior of a quantum fluid in reduced dimensions continues to elicit considerable research activity, in part motivated by
 recent experimental advances, allowing one to investigate, e.g., superfluid helium  films or cold-atom assemblies in novel, yet unexplored 
 settings. 
 \\ \indent
 The superfluid transition of a Bose fluid in three dimensions (3D) occurs at the critical temperature $T_c$, concomitantly with the onset of Bose-Einstein condensation, namely, the appearance of off-diagonal long-range order (ODLRO) \cite{Leggett2006,Kora2020b}. 
By contrast, in two dimensions (2D) the superfluid phase displays no true ODLRO at any finite temperature, but rather
 a slow (power-law) decay of spatial correlations.
The superfluid transition in 2D is theoretically understood within the Berezinskii-Kosterlitz-Thouless (BKT) general framework \cite{Berezinskii1972,Kosterlitz1972,Kosterlitz1973}; the characteristic fingerprint of 
the BKT  transition is the so-called ``universal jump'' of the superfluid fraction $\rho_s (T)$ as a function of 
temperature, from zero to a finite value as $T_c$ is approached from above \cite{Nelson1977,Jose1977,Ohta1979,Weber1988}. \\ \indent
Yet another paradigm change takes place if the system is confined to just one dimension
 (1D), for in that case a comprehensive description of its low-lying excitations 
and its ensuing thermodynamic properties is provided by the Tomonaga-Luttinger liquid    
theory \cite{Haldane1981}. While, strictly speaking, no superfluid phase exists in 1D in the thermodynamic 
limit (i.e., $L\to\infty$, $L$ being the system size), one can still meaningfully speak of ``superfluidity'' of a 1D
system as a well-understood and characterized finite-size effect, i.e., $\rho_s (L,T)$ is a universal function of $LT$
\cite{Haldane1981,Delmaestro2010,Delmaestro2010b}. It should also be noted that, although no superfluid (i.e., indefinitely long-lived) current can in principle be sustained in 1D, nonetheless the physical mechanism that leads to current decay in 1D, namely, phase slips \cite{Little1967,Langer1968,McCumber1970,Zaikin1997,Freire1997}, can be strongly suppressed at low temperature, 
to the point where there may be no practical experimental difference between a current-carrying state in 1D and a
 3D superfluid \cite{Kagan2000}.  Moreover, there exist theoretical scenarios in which 3D superflow could be 
 established in a network of interconnected quasi-1D channels \cite{Shevchenko1988,Boninsegni2007}.
\\ \indent 
Experimental verification of the BKT transition has been achieved in a variety of physical settings, including superfluid ($^4$He) \cite{Bishop1978,Agnolet1989,Csathy1998,Boninsegni1999,VanCleve2008,Kosterlitz2020} and superconducting \cite{Epstein1981} thin films, 
Josephson junction arrays \cite{Resnick1981}, and, relatively more recently, cold-atom assemblies \cite{Hadzibabic2006,Desbuquois2012,Fletcher2015,Sunami2022}. 
In order to observe Luttinger liquid behavior, several experimental avenues have been considered to confine quantum fluids such as $^4$He in (quasi) 1D. 
In particular, the adsorption of helium gas inside elongated cavities of nanometer-size diameter, such as those that exist in a variety of porous glasses \cite{Sokol1996,Dimeo1998,Plantevin2001,Anderson2002,Toda2007,Prisk2013}, or nanoholes in Si$_3$N$_4$ membranes \cite{Savard2011}, 
as well as carbon nanostructures \cite{Teizer1999,Ohba2016}, has been vigorously pursued, seen as it was as the most promising approach. More recently, however, interesting alternatives have emerged, such as self-assembled chains of atoms on surfaces \cite{Blumenstein2011} and cold atoms \cite{Kinoshita2004,Yang2017,Cedergren2017}.
\\ \indent
The remarkable degree of control that has been attained on many of the relevant systems that have been investigated  allows one to pose
 fundamental theoretical questions on the physics of superfluids in reduced dimensions, making predictions for which actual experimental
  verification may be feasible. One such question is whether it is possible, by tweaking an external parameter, to change the effective 
  dimensionality of a superfluid and observe the ensuing change in the behavior of the system, described by the  above-mentioned, 
  different theoretical frameworks \cite{Lammers2016}.
Some of these issues have already been explored in the context of dipolar assemblies of cold atoms or molecules, which can 
form 3D parallel stripes (elongated droplets in finite systems) \cite{Wenzel2017,Biagioni2022} whose collective behavior can
 mimic that of a 2D cluster crystal \cite{Boninsegni2012,Kora2019}. \\ \indent 
But even if interactions among the constituent particles are isotropic, one can imagine inducing a dimensional crossover by superimposing, 
e.g., to a quasi-2D Bose gas,  an external modulating potential of variable amplitude {\em along a specific direction}. In this setup, which is 
well within the reach of current experimental cold-atom technology \cite{Kinoshita2004,Meinert2015,Boeris2016}, one should observe 
the breakdown of the system into nearly independent, quasi-1D stripes (or ``tubes''), for sufficiently large amplitude of the external potential, conceivably accompanied by a change in the physical behavior of the system, reflecting an effective change of dimensionality, from 2D to 1D.
 This behavior would   allow, for instance, by means of   pertinent modulating potentials, to mimic  
 quasi 1D systems with nontrivial topology such as, for instance, junctions and/or networks of  1D channels 
 \cite{Moraal1976,Stan2000,Boninsegni2001,Giamarchi2004,Kalinay2014}  or 
 to realize in a tunable and controlled way the physics associated to the topological Kondo effect
\cite{Beri2012,Beri2013,Altland2013,Giuliano2020t,Giuliano2020tb,Giuliano2022,Buccheri2022}.
\\ \indent
With the aim of characterizing such a possible dimensional crossover, we investigate this scenario theoretically  within the framework of the 
classical $|\psi|^4$ lattice field theory. The reason for this choice is that, despite the obviously oversimplified description that this model 
provides of the system of interest, it nonetheless features all the physical aspects that we wish to explore; i.e., it displays a
 BKT superfluid transition while allowing for an externally induced density modulation, expressed through the use of a locally varying 
 chemical potential. 
It also lends itself to a semianalytical analysis, which we then validate quantitatively by means of large-scale, numerical simulations.
\\ \indent
Our main finding is that the uniaxial external modulation induces {\em no} dimensional crossover for any finite value of the amplitude 
of the modulation. Rather, as the system progressively forms quasi-1D parallel stripes in the direction perpendicular to that of the 
modulation, its physical behavior approaches that of the classical anisotropic $x$-$y$ model, i.e., with different coupling along the 
two directions. In particular, increasing the amplitude of the modulation has the effect of suppressing the superfluid transition temperature 
$T_c$, while the anisotropy of the superfluid response can be interpreted as a change of length scale in one of the two directions.
\\ \indent
The remainder of this paper is organized as follows: in Sec. \ref{modham} we introduce the model and discuss the main issue of interest,
 as well as the different investigative methodologies adopted in this work. In Sec. \ref{Sec_analytical}, we show that model (\ref{psi4}) becomes effectively equivalent to an anisotropic $x$-$y$ model in the limit of large modulation amplitude, and we 
 obtain analytical predictions concerning the superfluid transition. In Sec. \ref{antr} 
we  assess our analytical predictions against the results of our numerical (Monte Carlo) simulations. We offer our discussion and conclusions in Sec. \ref{concl}, while in the Appendix we 
 provide the mathematical details of the mapping between the $|\psi|^4$ model and the anisotropic $x$-$y$ model.

\section{Model}
\label{modham}
The classical $|\psi|^4$ field theory is defined by the following Hamiltonian
\begin{equation}
    \label{psi4}
    H = -t \sum_{\langle {\bf r r^\prime}\rangle}(\psi_{\bf r}\psi^\star_{\bf r^\prime}+\psi^\star_{\bf r}\psi_{\bf r^\prime})+
    \sum_{\bf r}\biggl (\frac{U}{2} n_{\bf r}^2-\mu_{\bf r} n_{\bf r} \biggr ) \: \: . 
\end{equation}
We assume a square lattice of $L\times L$ sites ($L$ even), with periodic boundary conditions in both directions; the position of 
a generic lattice site is ${\bf r}\equiv(l_x,l_y)$, with $l_x$ and $l_y$ being integers, $1 \le l_{x(y)}\le L $. The (first) second sum runs over all 
(pairs of nearest-neighboring) sites, $\psi_{\bf r}$ is a complex-valued field defined at site {\bf r}, and $n_{\bf r}=|\psi_{\bf r}|^2$ is 
the corresponding density  of particles. The parameter $t$ is a particle-hopping energy, which we take as our energy unit and set equal to 1. 
$U$ (assumed positive in this work) is the characteristic energy of interaction of particles occupying the same site, while  
$\mu_{\bf r}$ is a (site-dependent) chemical potential, which we assume to be of the following form
\begin{equation}
    \mu_{\bf r} = V_0 + V_1 \ {\rm cos}\biggl ( \frac {2\pi m l_y}{L}\biggr )
    \label{expot} \;\; . 
\end{equation}
\noindent
$\mu_{\bf r}$ accounts for an external potential, which is taken to be along the $y$ direction and has amplitude $V_1$.  $m$ is an
 integer number ranging from 1 to $L$  and commensurate with $L$, so that the modulation takes place 
 over a period of $N=L/m$. 
 \\ \indent 
Equation (\ref{psi4}) is the classical limit of the well-known Bose Hubbard model, approached when the 
average occupation number $\langle n_{\bf r}\rangle \gg  1$. 
In the absence of an external potential (i.e., with $V_1=0$), and with $V_0=U$, Eq. (\ref{psi4}) reduces to the well-known
$x$-$y$ model, in the strong coupling (i.e., $U\to\infty$) limit. In 2D, model (\ref{psi4}) displays a BKT superfluid transition, 
the role of the superfluid response being played by the classical helicity modulus \cite{Fisher1973}. It constitutes a suitable
 {\em minimal} model to gain insight into the physics of interest here, since we aim at determining whether a change in
  the effective dimensionality of the system occurs, for a finite value of the modulation amplitude. Such a change ought 
  to be mirrored in the critical properties of the system, which in turn reflect its behavior over long distances, 
  largely insensitive on whether the underlying field theory is formulated in the continuum or on a lattice 
   or whether it is quantum or classical in character. 
   \\ \indent
It is worth mentioning that the effect of an anisotropic hopping parameter, including the case of spatial modulation in one direction, has been studied in the context of the Bose-Hubbard model \cite{You2012}. In the model considered in this work, on the other hand, the anisotropy of the physical behavior, including a possible dimensional crossover, arises exclusively from the imposition of an external potential.
The advantage of utilizing Eq. (\ref{psi4}) as a starting point is that it allows one to establish some basic physical conclusions analytically and test them reliably by means of large-scale numerical (Monte Carlo) simulations.

\section{Anisotropic $x$-$y$ model description of the classical $|\psi|^4$ theory}\label{Sec_analytical}

Model (\ref{psi4}) can be shown to be effectively equivalent to an anisotropic $x$-$y$ model.
We begin by re-expressing the $|\psi|^4$ Hamiltonian using the ``polar coordinate'' representation for 
$\psi_{\bf r}$ given by $\psi_{\bf r}=\sqrt{n_{\bf r}}\ e^{i\theta_{\bf r}}$, i.e.,
\begin{equation}
    H = -\sum_{\langle {\bf r r^\prime}\rangle}\ t \sqrt{n_{\bf r}n_{\bf r^\prime}}  {\rm cos}(\theta_{\bf r}-\theta_{\bf r^\prime})
    + \sum_{\bf r}\biggl (\frac{U}{2} n_{\bf r}^2-\mu_{\bf r} n_{\bf r} \biggr )
    \: .
    \label{cores.5}
  \end{equation} 
 For $U$ large and  $V_0=U$ one may rely on a saddle-point approximation of 
 the right-hand side of Eq. (\ref{cores.5}). To do so, one sets $n_{\bf r} = \bar n_{\bf r} + 
 \delta n_{\bf r}$, with  $\bar n_{\bf r}$ being the saddle-point solution for $n_{\bf r}$. 
 Moreover, fluctuations in the phase differences $\theta_{\bf r}- \theta_{{\bf r^\prime}}$ are typically 
 assumed to be of order $(\delta n_{\bf r})^2$ \cite{Wallin1994}.  Taking that into account, we conclude that
 $\cos(\theta_{\bf r}-\theta_{\bf r^\prime})\approx 1 + {\cal O}((\delta n_{\bf r})^2)$. Therefore,   neglecting
  the coupling between $\delta n_{\bf r}$ and the fluctuations of 
$\theta_{\bf r}$ up to  second-order in the fluctuations, we obtain 
   
 \begin{eqnarray}
&& -\sum_{\langle{\bf r r^\prime }\rangle}\ t \sqrt{n_{\bf r}n_{\bf r^\prime}}\cos(\theta_{\bf r}-\theta_{\bf r^\prime})
 \approx   - \sum_{\langle{\bf r r^\prime}\rangle   }\ t \sqrt{n_{\bf r}n_{\bf r^\prime}} \nonumber \\
&&  +\sum_{\langle{\bf r r^\prime}\rangle}\ t\sqrt{\bar{n}_{\bf r}
  \bar{n}_{\bf r^\prime}} [1-\cos(\theta_{\bf r}-\theta_{\bf r^\prime})]
  \:\:.
  \label{cores.x1}
  \end{eqnarray}
  \noindent
  Inserting Eq. (\ref{cores.x1}) into Eq. (\ref{cores.5}) and equating to 0 the term that is linear in 
  $\delta n_{\bf r}$, we recover the saddle-point equations for  $\bar{n}_{\bf r}$. These are given by

\begin{eqnarray}
 &&   t 
\{         \sqrt{ \bar{n}_{(l_x+1,l_y) }}  +    \sqrt{ \bar{n}_{(l_x-1,l_y)} }  + \nonumber \\ 
  &&           \sqrt{ \bar{n}_{(l_x ,l_y+1) }}  +   \sqrt{ \bar{n}_{(l_x ,l_y-1)} }  \} \nonumber \\
&&  =  \sqrt{\bar{n}_{(l_x,l_y)}  } \{ U \bar{n}_{(l_x,l_y)}  - \mu_{(l_x,l_y)}  \} 
\:\:\:\:,
\label{cores.10}
\end{eqnarray}
\noindent
with the additional constraint that $\forall {\bf r}$ one has $\bar{n}_{\bf r} \geq 0$. When 
$t=0$, Eq. (\ref{cores.10}) reduces to the ``local density approximation'' solution,
$\bar{n}_{(l_x,l_y)}=\mu_{(l_x,l_y)}/U$ if $\mu_{(l_x,l_y)}>0$, and $=0$ otherwise.
A finite $t$, instead, 
implies a finite $\bar{n}_{(l_x,l_y)}$ over each lattice site, even  for $|V_1| >  |V_0|$. 

The ``leftover'' term at the right-hand side of Eq. (\ref{cores.x1}), which does not depend on 
$\delta n_{\bf r}$, eventually provides  the effective   Hamiltonian 
 describing  the phase fluctuations of the $|\psi|^4$ model 
  (that are the relevant, low-lying degrees of freedom close to the BKT  phase transition 
  \cite{Wallin1994}).  Substituting 
each $ \bar{n}_{\bf r}$ at the right-hand side of Eq. (\ref{cores.x1}) with the corresponding saddle-point
solution  of Eq.  (\ref{cores.10}),  we conclude that the phase fluctuations are described by 
   the modulated $x$-$y$ Hamiltonian $H_{x-y}^{\rm mod}$, given by 
  
    \begin{eqnarray}
 H_{x-y}^{\rm mod}       = 
 &-& 2 \sum_{\bf r}  \{  J_{\bf r}^x \cos [         \theta_{   (l_x+1,l_y)}   - \theta_{(l_x , l_y)}   ]  \nonumber \\
 &+&  J^y_{\bf r}  \cos [         \theta_{(l_x , l_y+1)}  - \theta_{(l_x,l_y)}  ]          \} 
\:\:\:\:,
\label{mh.1}
\end{eqnarray}
\noindent
with $J_{{\bf r}}^x = t \sqrt{ \bar{n}_{(l_x,l_y)} \bar{n}_{  (l_x+1,l_y)}} $ and $J_{{\bf r}}^y = t \sqrt{ \bar{n}_{(l_x,l_y)}
\bar{n}_{  (l_x ,l_y + 1)}} $.  Given the periodic form of the uniaxial modulation
(\ref{expot}),   we
obtain that $J_{ (l_x,l_y+N)}^{x(y)} = J_{(l_x,l_y)}^{x(y)}$, with $N = L / m$ being the modulation
period. Moreover, since $\bar{n}_{(l_x,l_y)}$ is uniform along the $x$ direction, 
(that is, it is independent of $l_x$, just as $\mu_{(l_x,l_y)}$), we infer that both $J^x$ and $J^y$ are functions
of $l_y$ only. Finally, as we evidenced above, $\bar{n}_{\bf r}$ is 
finite over every lattice site,  which implies that $J^x_{\bf r}$ and $J_{\bf r}^y$ are
 different from 0 over every bond of the lattice. 
\\ \indent
Given the correspondence between $H$ in Eq. (\ref{psi4}) and  $H_{x-y}^{\rm mod}$, 
we refer to  this latter model Hamiltonian to compute
  the superfluid fractions in the two directions as a function of the temperature $T$,
$\rho_{s,x} ( T)$ and $\rho_{s,y}(T)$. Specifically  \cite{Ohta1979}, we ``twist'' 
 $\theta_{\bf r}\to \theta_{\bf r}+{\cal Q}_x \frac{l_x}{L}+{\cal Q}_y \frac{l_y}{L}$ and 
identify the superfluid fractions $\rho_{s,x} (T)$ and  $\rho_{s,y} (T)$ from the coefficients of the quadratic
(in ${\cal Q}_x$ and ${\cal Q}_y$)  contributions to the total free energy. In the low-temperature limit, we 
resort  to an ``improved'' Villain approximation \cite{Ohta1979}, i.e.,  we expand 
$\cos (\theta_{\bf r}-\theta_{{\bf r}^{'}})$ up to fourth-order in $\theta_{\bf r}-\theta_{{\bf r}^{'}}$. 
{Expanding up to fourth-order allows us to recover} the leading, low-$T$ contributions to $\rho_{s,x(y)}(T)$ and 
$\rho_{s,x(y)}^{(0)}(T)$, {\it without} accounting for the contributions from vortex excitations, 
which we introduce later on, within the renormalization group (RG) approach to the BKT phase transition.
\indent \\
In implementing the Villain approximation, one has to account suitably for the  ``superperiodicity'' induced by the modulation.
To do so, we write the Fourier mode expansion of $\theta_{(l_x,l_y)}$ and of $J_{l_y}^{x(y)}$ as

\begin{eqnarray}
\theta_{(l_x,l_y)} &=& \frac{1}{L^2}\sum_{\vec{k} \in {\cal B}_N} \sum_{\nu = 0}^{N-1} e^{i\vec{k}\cdot{\bf r} + \frac{2 \pi i \nu l_y}{N}} 
\: \theta_{\vec{k},\nu} \nonumber \\
J_{l_y}^{x(y)}&=&\frac{1}{N}\sum_{\nu=0}^{N-1} e^{\frac{2 \pi i \nu l_y}{N}} J^{x(y)} (\nu) \  \:\:\: ,
  \label{mxy.3}
 \end{eqnarray}
 \noindent
 with the reduced Brillouin zone ${\cal B}_N=[-\pi,\pi] \times \left[-\frac{\pi}{N},\frac{\pi}{N}\right]$. 
 To recover the large-scale, low-${\bf k}$ effective description of our system, we systematically
 integrate over the $\theta_{{\bf k},\nu}$-modes, with $\nu \neq 0$ so as to obtain 
 an effective Hamiltonian only involving the $\nu=0$ modes. In the Appendix   we describe in
 detail the whole derivation. As a result, we eventually obtain

 \begin{eqnarray}
 && H_{{\rm Eff},{\rm mod}}^{\rm Vil} [\vec{\cal Q}] = 
 \frac{1}{2 L^2} \sum_{{\bf k} \in {\cal B}_N} \Delta ({\bf k}) | \theta_{{\bf k},0}|^2 \nonumber \\
 && + \frac{ [{\cal Q}_x^2 {\cal J}^x ( T ) + {\cal Q}_y^2 {\cal J}^y ( T) ]}{2 N}  
    \;\;\; . 
\label{mxz.1}
\end{eqnarray}
\noindent
In the Appendix  we show that ${\cal J}^{x(y)} (T) = {\cal J}^{x(y)}_0-T{\cal J}^{x(y)}_1$ and 
we provide the explicit formulas for ${\cal J}^{x(y)}_0$ and for ${\cal J}_1^{x(y)}$.  Therefore, from  Eq. (\ref{mxz.1}) we  determine the 
(``bare,'' that is, undressed by vortices) superfluid fractions along the two directions,  according to 

\begin{eqnarray}
\rho_{s,x}^{(0)}  (T)&=&\frac{{\cal J}^x ( T)}{{\cal J}^x ( 0)} = 1 - \frac{T}{\delta_x}\;\; ,  \nonumber \\
\rho_{s,y}^{(0)} (T)&=&\frac{{\cal J}^y ( T)}{{\cal J}^y (  0)} = 1 - \frac{T}{\delta_y}
\;\; ,
\label{mxy.15}
\end{eqnarray}
\noindent
with $\delta_{x(y)} = {\cal J}_0^{x(y)}/{\cal J}_1^{x(y)}$.   In Eq. (\ref{mxz.2}) we provide the explicit 
formula for the kernel $\Delta ({\bf k})$. By expanding $\Delta ({\bf k})$
up to second order in ${\bf k}$, we obtain 

\begin{eqnarray}\nonumber
 H_{{\rm Eff},{\rm mod}}^{\rm Vil}= H_{{\rm Eff},{\rm mod}}^{\rm Vil} [\vec{\cal Q} =  {\bf 0}] \\
 \approx \frac{1}{2L^2} \sum_{{\bf k}} \{ {\cal J}^x (T) k_x^2 + 
 {\cal J}^y (T) k_y^2 \}| \theta_{{\bf k},0}|^2 
 \:\:\: .
 \label{mxz.3}
 \end{eqnarray}
 \noindent 
 The right-hand side of Eq. (\ref{mxz.3}) corresponds to the long-wavelength expansion of the Hamiltonian of 
 a uniform,   anisotropic $x$-$y$ model with coupling strengths in the two directions respectively given by 
 ${\cal J}^x(T)$ and ${\cal J}^y (T)$. Therefore, in the following we employ this latter model    to account for the effect of the vortices
 on the superfluid fractions.  
 \\ \indent
{\em The BKT superfluid transition}. 
In the general framework of the $x$-$y$ model it is well established 
that, on taking into account  vortex 
excitations, the ``renormalized'' superfluid fractions   $\rho_{s , x(y)} $  acquire an explicit dependence 
on the running scale $\lambda$ (eventually identified with the system size)
\cite{Jose1977,Ohta1979,Itzykson1989}.  Denoting with $y ( T , \lambda )$, with   $\rho_x ( T , \lambda) $, 
and  with $\rho_y (T , \lambda )$, respectively, the scale-dependent single-vortex fugacity and the superfluid fractions, 
their scaling with $\lambda$ is described by the (anisotropic)  RG equations given by  
\cite{Jose1977,Itzykson1989,You2012}

\begin{eqnarray}
\frac{d y(T ,\lambda )}{d \ln \lambda}&=& \left[2 - \frac{ \pi   {\cal J}}{T} 
 \sqrt{\rho_{s,x}(T , \lambda)\rho_{s,y}(T,\lambda)}\right ]y(T,\lambda) \; ,  \nonumber \\
\frac{d \rho_{s,x}(T,\lambda)}{d \ln \lambda}&=&-\frac{  2\pi^3 {\cal J}}{T}  y^2(T,\lambda)\sqrt{[\rho_{s,x}(T,\lambda)]^3\rho_{s,y}(T,\lambda)}
\; , 
\nonumber \\
\frac{d \rho_{s,y}(T,\lambda)}{d \ln \lambda}&=&-\frac{  2\pi^3 {\cal J}}{T}  y^2(T,\lambda)\sqrt{[\rho_{s,y}(T,\lambda)]^3\rho_{s,x}(T,\lambda)} \; ,
\nonumber \\
\label{vil.18}
\end{eqnarray}
\noindent
with ${\cal J}=\sqrt{{\cal J}^x (0) {\cal J}^y (0) }$. The superfluid fractions in the thermodynamic limit are recovered from the solutions of 
Eqs. (\ref{vil.18}) at given $\lambda$ and $T$, determined by using the bare superfluid fractions in 
Eqs. (\ref{mxy.15}) as initial values of the parameters at the reference scale, by eventually taking the $\lambda \to \infty$-limit. 
The single-vortex fugacity at the reference scale, $y^{(0)} (T)$,  can be estimated using a pertinent extension to the anisotropic
model of the results of Ref. \cite{Ohta1979}, which is  
$y^{(0)} (T) \approx \exp \left[ - \frac{\pi^2 {\cal J} }{2 T} \right]$. 
Over a finite-size ($L^2$) lattice, we recover the finite-size superfluid fractions 
 $\rho_{s , x(y)} ( T , L )$ by stopping the  RG flow  determined by 
Eqs. (\ref{vil.18})  at $\lambda = L$.  

To solve Eqs. (\ref{vil.18}), we note that they imply that the dimensionless quantity 
${\cal K} ( T ) = \rho_{s,y} (T, \lambda ) / \rho_{s,x}  (T, \lambda )$ is constant along the RG trajectories; that is, 
it is independent of $\lambda$.  Accordingly, we set

\begin{eqnarray}
\rho_{s,x} ( T , \lambda ) &=&  \rho_s ( T , \lambda )\sqrt{{\cal K} (T)} \;\;\; , \nonumber \\
\rho_{s,y} ( T , \lambda ) &=&  \rho_s ( T , \lambda )/ \sqrt{{\cal K} (T)} 
\:\:\: . 
\label{mz.1}
\end{eqnarray}
\noindent
In terms of $y(T, \lambda )$ and $\rho_s ( T , \lambda )$, the system (\ref{vil.18}) reduces to

\begin{eqnarray}
\frac{d y(T ,\lambda )}{d \ln \lambda}&=& \left[2 - \frac{\pi {\cal J}}{T}  \rho_{s}(T , \lambda) \right]y(T,\lambda)\;\; ,  \nonumber \\
\frac{d \rho_{s}(T,\lambda)}{d \ln \lambda}&=&- \frac{2\pi^3{\cal J}}{T}  y^2(T,\lambda) \rho_{s}(T,\lambda)^2
\:\:.
\label{vil.19}
\end{eqnarray}
\noindent
Equations (\ref{vil.19})  correspond  to the familiar set of the BKT RG  equations for the running parameters in 
  the isotropic $x-y$ model \cite{Itzykson1989}.   It is, therefore, immediate to infer they  
  imply that the critical temperature  $T_c$ satisfies the equation \cite{Itzykson1989} 

\beq
2 \pi y^{(0)}  ( T_c  ) 
+ 2 -\frac{\pi{\cal J}}{  T_c} \rho_s^{(0)} (T_c) = 0 
\:\:\:\: , 
\label{ex.3}
\eneq
\noindent
with $ \rho_s^{(0)} (T) = \sqrt{ \rho_{s,x}^{(0)} (T) \rho_{s,y}^{(0)} (T)}$. 
(Roughly speaking, Eq. (\ref{ex.3}) implies a scaling of $T_c$
 with ${\cal J}$, as it is   typical of the anisotropic $x$-$y$ model \cite{Williams2006}).
 
\begin{figure}
 \center
\includegraphics*[width=1 \linewidth]{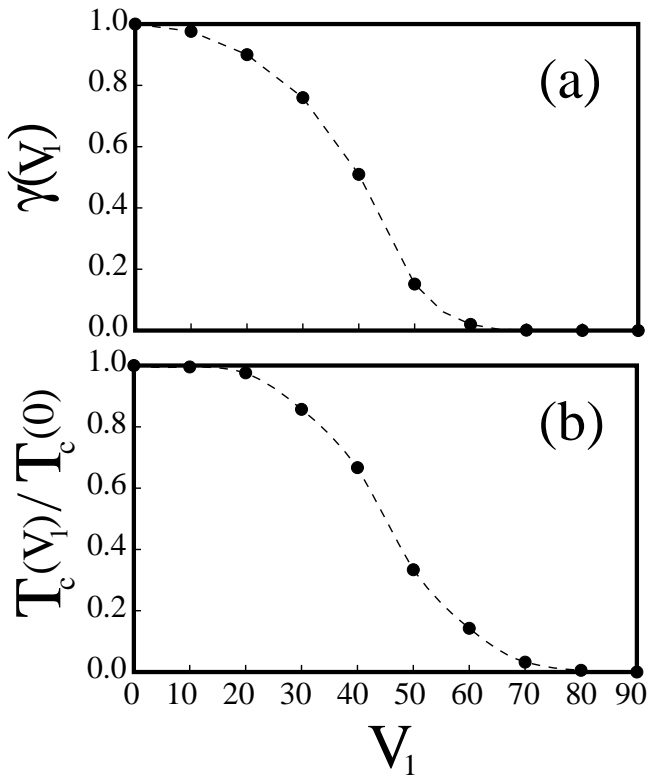}
\caption{(a)  Effective anisotropy $\gamma(V_1) \equiv {\cal J}_y(0)/{\cal J}_x (0)$  induced in the equivalent $x$-$y$ Hamiltonian $H_{\rm Eff,mod}^{\rm Vil}$ 
by the external modulation.
(b)  Critical temperature for the BKT phase transition as a function of $V_1$, $T_c(V_1)$, normalized to the critical temperature 
in the absence of modulation, $T_c(0)$.  In both panels, the interpolating dashed line is a guide to the eye.    } 
\label{anis}
\end{figure}

 Finally, we   recover  the ``anisotropic'' universal jump condition for the superfluid 
 fractions, consistent with ${\cal K} ( T )$ being invariant along 
the RG trajectories, given by    
  
 \begin{eqnarray} 
&& \lim_{T \to T_c^-} \rho_{s,x}  (  T   )  = \frac{ 2  T_c    }{ \pi  {\cal J} \sqrt{{\cal K}(T_c)} } \; ,  \label{appe.14} \\
&& \lim_{ T \to T_c^+ } \rho_{s,x}  ( T   )  =  0 
\:\:\:\: ,  \nonumber
\end{eqnarray}
\noindent
and

 \begin{eqnarray} 
&& \lim_{T \to T_c^-} \rho_{s,y}  (  T   )  = \frac{ 2  T_c   \sqrt{{\cal K} (T_c )}  }{ \pi   {\cal J} }\; ,   \label{appe.15} \\
&& \lim_{ T \to T_c^+ } \rho_{s,y}  ( T   )  =  0 
\:\:\:\: ,  \nonumber
\end{eqnarray}
\noindent
with $\rho_{s,x(y)} (T) = \lim_{\lambda \to \infty} \rho_{s,x(y)} (T , \lambda )$. 
\\ \indent
For a finite system size $L$,  Eqs.  (\ref{vil.18}) 
predict   a downturn in both $\rho_{s,x}(T,L)$ and $\rho_{s,y}(T,L)$ as a function of $T$, centered over a certain ``finite-size
critical temperature'' $T_c (L)$ (which is the same for both the superfluid fractions). The larger $L$ is, the sharper the downturn 
in the superfluid fractions is. In the thermodynamic limit $L\to \infty$, the downturn evolves into the sharp ``universal critical 
jump'': the fingerprint of the BKT phase transition in a two-dimensional system  \cite{Nelson1977,Jose1977,Ohta1979}. 
\\ \indent
The uniaxial modulation induces an effective anisotropy, as illustrated in 
Fig. \ref{anis} (a), where the ratio $\gamma(V_1) \equiv {\cal J}_y(0)/{\cal J}_x (0)$, computed 
based on Eqs. (\ref{mxy.9}) and  (\ref{mxy.14}), is shown for the value of the model parameters used here (see above). There is a monotonic decrease, the system remaining two-dimensional for arbitrarily large values of $V_1$.   It is worth stressing that, modulating the hopping rather than the potential \cite{You2012}  would possibly lead to a similar effective description of the scaling of the superfluid fractions. 
\\ \indent
 Figure \ref{anis} (b)  also shows the computed critical temperature for the BKT phase transition as a function of $V_1$, normalized to the critical temperature 
in the absence of modulation, $T_c(V_1) /T_c(0)$, as a function of $V_1$. 
As one might intuitively expect, the quantities shown in Figs. \ref{anis} (a) and \ref{anis} (b) behave similarly as a function of $V_1$.
 Indeed, in the  limit $ |V_1/V_0| \ll 1$, a perturbative calculation based on the formalism of appendix \ref{appe} shows that both quantities decrease quadratically with $V_1$, while in the opposite limit the numerical results indicate a change of convexity; i.e., both quantities approach zero asymptotically.

\begin{figure}
 \center
\includegraphics*[width=1 \linewidth]{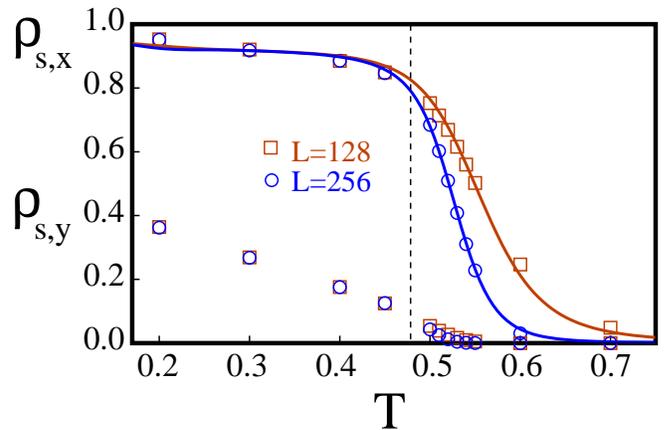}
\caption{Superfluid responses $\rho_{s,x} ( T , L )$  and $\rho_{s,y} (T , L )$ of model (\ref{psi4}), computed by means of  the
Monte Carlo approach as a function of $T$ for  
$V_0 = U = 40$, $t=1$, and modulation strength $V_1=40$, for different system sizes.   
 The dashed vertical line marks the location of the transition, i.e., $T=T_c$. Also shown are fitting curves obtained 
 as explained in the text.} 
\label{40_all}
\end{figure}

{\section{numerical results}
\label{antr}}In order to obtain an unbiased, numerical check of our predictions,  
we performed Monte Carlo numerical simulations of the lattice field theory (\ref{psi4}), specifically computing the superfluid responses $\rho_{s,x} (T,L)$ and $\rho_{s,y} (T,L)$ 
as a function of $T$ for various system sizes. We used the classical worm algorithm, in its standard lattice implementation described, for instance, in Ref. \cite{Prokofev2001}. In particular, the superfluid fraction is estimated by means of the well-known winding number estimator.\\ \indent
We henceforth take $t$ as our energy unit, and set $V_0=U=40$; i.e., we work in the strong-coupling limit of the theory, in which
Eq. (\ref{psi4}) approaches the isotropic $x$-$y$ model in the absence of external modulation. 
For definiteness, but without any loss of generality, we set the period of the modulation of the external potential 
$N=8$ lattice sites. 
\\ \indent
Figure \ref{40_all} shows Monte Carlo results for $\rho_{s,x} (T,L)$ and $\rho_{s,y}(T,L)$, computed for two different system sizes, namely, $L=128$ and $L=256$, for a value of the amplitude of the modulating external potential $V_1=40$.
 The downturn in both $\rho_{s,x}(T,L)$ and   $\rho_{s,y}(T,L)$ at a temperature of
 $T_c \sim 0.48$ is clear, although it is less evident in $\rho_{s,y} (T,L)$, due to the 
 anisotropy-induced reduction of the superfluid fraction in the direction of the modulation \cite{Williams2006}. As expected, the transition becomes increasingly sharp as $L$ grows; despite the presence of the modulating field,  the evidence of a 
 BKT phase transition in the planar model seems clear. Obviously, however, this assertion must be verified by carrying out finite-size scaling analysis.

\begin{figure}
 \center
\includegraphics*[width=1 \linewidth]{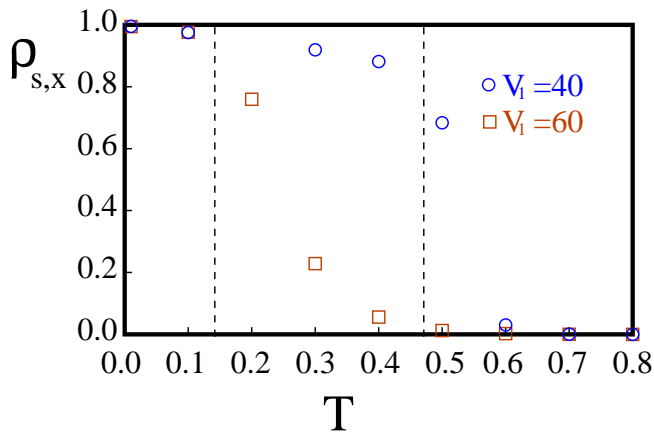}
\caption{ Superfluid response $\rho_{s,x}(T,L)$ of the model (\ref{psi4}) as a function of $T$, computed by Monte Carlo simulations for a square lattice with $L^2 =256^2$ sites. The amplitudes of the modulation are $V_1=40$  (circles) and $V_1=60$ (squares), while the values of all other model parameters are specified in the text. 
The   dashed, vertical lines mark  the approximate locations of 
the two critical temperatures.  } 
\label{ccritt}
\end{figure}

\begin{figure}
 \center
\includegraphics*[width=1 \linewidth]{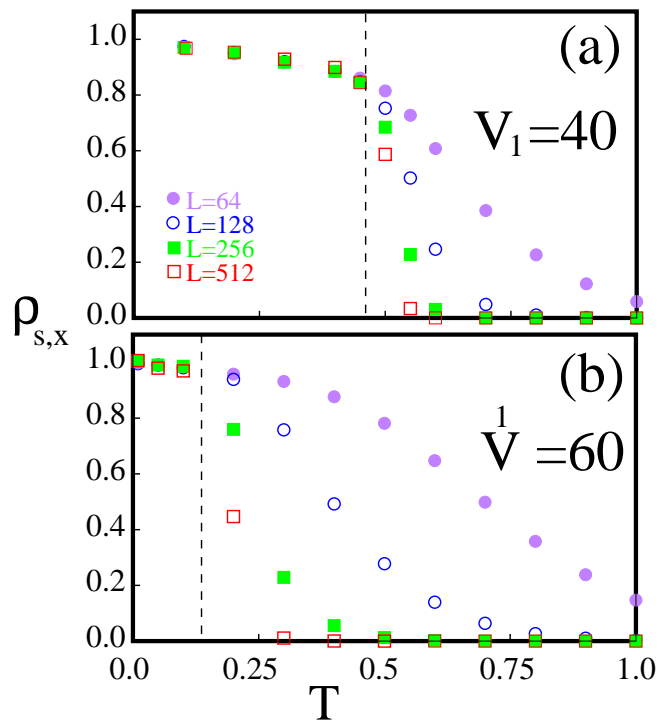}
\caption{Superfluid response $\rho_{s,x}(T,L)$ of the model (\ref{psi4}) as a function of $T$, computed by Monte Carlo simulations for a square lattice for  $L=64$ (purple solid dots), $L=128$ (blue empty dots), $L=256$ (green solid squares), and $L=512$
(red empty squares),   and for $V_1=40$ (a) and $V_1=60$ (b).  The dashed 
vertical lines mark the (approximate) location of the critical temperature in the two cases.  } 
\label{scaling}
\end{figure}

\indent 
On integrating the RG equations (\ref{vil.18})  up to $\lambda= L$ for different  values of $T$, one can obtain fitting curves for 
 $\rho_{s,x}(T,L)$ and $\rho_{s,y}(T,L)$. To do this analytically, one needs to know how
 the system parameters at the reference scale depend on the temperature.
Equations (\ref{mxy.15}) and the expressions for $\delta_x$ and $\delta_y$ in the Appendix 
rely upon approximations that are strictly speaking only valid in the $T\to 0$ limit 
and thus are not expected to hold quantitatively near $T_c$. For this 
reason, we fitted the Monte Carlo results with
the curves described in Sec. \ref{Sec_analytical}, using $\delta_x, \delta_y$, and ${\cal J}$ as adjustable fitting parameters.
The excellent fit to the numerical data obtained in this way (shown in Fig. \ref{40_all}) represents strong evidence to the effect that   
the superfluid properties of model (\ref{psi4}) are the same as  those of the (anisotropic) $x$-$y$ model 
\footnote{Similar plots can be drawn for $\rho_{s,y}$ as well, but the fitting procedure is rendered complicated by the small values of 
$\rho_{s,y}$ around $T_c$, making the agreement with numerical data less impressive than for the $x$-part.}.
\\ \indent
Within the framework of the anisotropic $x$-$y$ model, one expects a reduction of  $T_c$ with increasing anisotropy, consistent with 
Eq. (\ref{ex.3}). 
In Fig. \ref{ccritt}, we show  $\rho_{s,x}(T,L)$  as a function of $T$. The reduction of $T_c$ on increasing $V_1$ (that is, 
the anisotropy in the effective $x$-$y$ Hamiltonian) is apparent (in the figure we mark with dashed vertical lines the approximate locations of 
the two critical temperatures) and is also roughly consistent with the 
results for the anisotropy {and for the critical temperature} in Fig. \ref{anis} and with the implication of Eq. (\ref{ex.3}).
\\ \indent
In our view, these results provide robust numerical confirmation of the theory described in Sec. \ref{Sec_analytical}, 
namely,  that the superfluid behavior of the 
$|\psi|^4$ theory in the presence of a uniaxial modulation reduces to that of the two-dimensional  anisotropic $x$-$y$ model. Accordingly,  increasing 
the strength of the modulation simply enhances the anisotropy, thus pushing  the BKT phase transition to lower values of $T_c$ but without 
determining any dimensional crossover in the system.   There is always a finite, though small,  $T_c$ at which the system undergoes the BKT phase transition from 
the superfluid to the disordered phase. 
\\ \indent
{To strengthen our conclusion that $T_c$ remains finite in the $V_1\to\infty$ limit, in Fig. \ref{scaling} we show our numerical results for $\rho_{s,x} (T,L)$ as a function of $T$ for increasing values of $L$, from 
$L=64$ till $L=512$, for $t=1$, $U=V_0=40$, and $V_1=40$ (a), and $V_1=60$ (b). In both cases we 
recognize the typical scaling of the superfluid fractions in the anisotropic $x-y$ model, with $T_c$ finite and consistent 
with the fitted data for $T_c$ as a function of $V_1$ in Fig. \ref{anis}. }
 
\bigskip

\section{Discussion and Conclusions}
\label{concl}
In this paper, we investigated the effects of a uniaxial external modulation over a two-dimensional superfluid.
We described the superfluid at finite temperature by means of the classical $|\psi|^4$ model over a square lattice. 
Adding the modulation on top of the well-established mapping between the $|\psi|^4$ model and the $x$-$y$ model, we 
derived a version of the latter model Hamiltonian with modulated parameters, which allowed us to spell out the 
effects of progressively increasing the potential modulation strength $V_1$.
\\ \indent
We show that despite the tendency of the system to develop quasi-1D stripes perpendicular to  
the direction of the modulation, at any $V_1$ the superfluid phase transition is well captured by the classical 
anisotropic $x$-$y$ model, to which the modulated model reduces in the long-wavelength, low-energy limit. In 
particular, the main effect of increasing $V_1$ is that of enhancing the anisotropy of the 
effective $x$-$y$ Hamiltonian and, correspondingly, pushing $T_c$ toward lower (though finite) values 
\cite{Yamashita2009,Yang2020,Nava2022}. 

Due to the wide applicability of our minimal model to describing the superfluid phase transition in 
planar, interacting bosonic systems, we infer that, as a general result, an external uniaxial modulation fails to induce a 2D to 1D dimensional crossover in such a system. The  good agreement between the analytical prediction and the 
numerical Monte Carlo data witnesses the reliability of our results, regardless of the various approximations we employed
along our derivation. In addition, the finite-size scaling analysis of the superfluid fractions  unambiguously shows that even for relatively large modulation amplitudes, the scaling behavior is that expected for a 2D system, which is completely different from  the 1D case \cite{Jose1977,Itzykson1989,You2012,Delmaestro2010,Delmaestro2010b,Nava2022}. 

Possible further extensions of our work  include, but are not limited to, considering the inclusion of disorder 
in the sample. It would be interesting to evidence whether the scenario we evidenced is affected by impurities. In this 
direction, given  the high level of control  reached in the technology of cold atom devices 
one may think, for instance, of engineering impurities ``ad hoc'', with tunable parameters, mimicking junctions of
quantum wires \cite{Chamon2003,Oshikawa2006,Hou2012,Giuliano2007,Giuliano2019,Kane2020,Guerci2021},
 or even network of junctions \cite{Medina2013}, with a high level of quantum coherence 
\cite{Novais2005,Giuliano2008} and a plethora of potential practical applications.  
 \vspace{0.3cm} 
 
{\bf Acknowledgements:} We thank Pasquale Sodano, 
 Andrea Trombettoni, and Nikolay Prokof'ev for insightful discussions. A.N. and D.G. 
  acknowledge   financial support  from Italy's MIUR  PRIN project  TOP-SPIN  (Grant No. PRIN 20177SL7HC). This work was also supported by the Natural Sciences and Engineering Research Council of Canada (NSERC). M.B. acknowledges the hospitality of the Universit\`a della Calabria, where part of this research work was carried out.
 
 \appendix
 
 \section{Derivation of Eqs.(\ref{mxz.1})}
 \label{appe}
 
 In this appendix we show that, once expressed in terms of the $\theta_{{\bf k},0}$, the 
 Hamiltonian $H_{{\rm Eff},{\rm mod}}^{\rm Vil} [\vec{\cal Q} ]$ takes the form in 
 Eq. (\ref{mxz.1}). 
 
We begin with the mode expansions in Eq. (\ref{mxy.3}). Denoting with 
  $H_{x-y}^{\rm mod} [ \vec{\cal Q}]$ the Hamiltonian in Eq. (\ref{mh.1}) at nonzero $\vec{\cal Q}$, 
  we approximate 
 
 \beq
 H_{x-y}^{\rm mod} \approx H_{{\rm Vil},2}^{\rm mod} [\vec{\cal Q} ]  +H_{{\rm Vil},4}^{\rm mod}  [\vec{\cal Q} ]  \;\; .
 \label{mmz.1}
 \eneq
 \noindent
The quadratic term in Eq. (\ref{mmz.1}) is given by 
 
  \begin{eqnarray}
 && H_{{\rm Vil},2}^{\rm mod} [\vec{\cal Q} ]  =\frac{1}{2 NL^2} \sum_{\nu,\nu'=0}^{N-1} \sum_{\vec{k} \in {\cal B}_N} \theta_{-\vec{k},-\nu}\theta_{\vec{k},\nu'} 
 {\cal D}_{\nu,\nu'} (\vec{k} ) \nonumber \\
 &&+\frac{ [{\cal Q}_x^2 J^x ( 0 ) + {\cal Q}_y^2 J^y (  0 ) ]}{2 N} \nonumber \\
 && + \frac{ {\cal Q}_y}{NL} \sum_{\nu=1}^{N-1} \theta_{{\bf 0},\nu} (e^{\frac{2 \pi i \nu}{N}}-1) J^y (-\nu) 
 \:\:\:\: ,
 \label{mxy.10}
 \end{eqnarray}
 \noindent
 with 
 
  \begin{eqnarray}
&&  {\cal D}_{\nu,\nu'}(\vec{k})=
 |1-e^{ik_x}|^2  {J}_x(\nu-\nu') \nonumber \\
 &&+(1-e^{-ik_y-\frac{2\pi i \nu}{N}} )
 (1-e^{ ik_y+\frac{2\pi i \nu'}{N}} )  {J}_y(\nu-\nu')
 \:\:\: .
 \label{mxy.11}
 \end{eqnarray} 
 As for the quartic term, we treat it within the mean-field approximation, 
 along the derivation of Ref. \cite{Ohta1979}. This implies decoupling quartic and cubic terms, 
 respectively, according to $\sum_{\bf r} J_{l_y}^{x} [\theta_{(l_x +1,l_y)}-\theta_{(l_x,l_y)}]^4 
 \to  \sum_{\bf r} J_{l_y}^x [\theta_{(l_x+1,l_y)}-\theta_{(l_x,l_y)}]^2 \langle    [\theta_{(l_x+1,l_y)}-\theta_{(l_x,l_y)}]^2 \rangle$,
 together with the analogous expression with $x\to y$,  
 and to $\sum_{\bf r} J_{l_y}^{y} [\theta_{(l_x ,l_y+1)}-\theta_{(l_x,l_y)}]^3 \to
 \sum_{\bf r} J_{l_y}^y [\theta_{(l_x,l_y+1)}-\theta_{(l_x,l_y+1)}]  \langle    [\theta_{(l_x ,l_y+1)}-\theta_{(l_x,l_y)}]^2 \rangle$, 
 with $\langle \ldots \rangle$ denoting the average with respect to the quadratic Hamiltonian  
 (\ref{mxy.10}).   Just as for the homogeneous, isotropic Hamiltonian, the contributions obtained in this way simply 
 amount to adding finite-$T$ corrections to $J^x(\nu)$ and $J^y(\nu)$, according to

 \begin{eqnarray}
&& J^x(\nu) \to \hat{J}^x ( \nu , T) =  J^x (\nu) - \frac{N T}{2 } \sum_{\nu_1 , \nu_2=0}^{N-1} J^x(\nu+\nu_1-\nu_2)  \nonumber \\
 && \times \frac{1}{L^2} \sum_{\vec{q} \in {\cal B}_N} 
 | 1 - e^{iq_x}|^2 [{\cal D} (\vec{q} )]^{-1}_{\nu_1,\nu_2}  \label{mxy.9}  \\
 && J^y(\nu) \to \hat{J}^y ( \nu , T) =  J^y (\nu) - \frac{N T}{2 } \sum_{\nu_1 , \nu_2=0}^{N-1} J^y(\nu+\nu_1-\nu_2)  \nonumber \\
 &&  \times\frac{1}{L^2} \sum_{\vec{q} \in {\cal B}_N} 
 ( 1 - e^{-iq_y - \frac{2\pi i \nu_1}{N}}) ( 1 - e^{iq_y + \frac{2\pi i \nu_2 }{N}})  [{\cal D} (\vec{q} )]^{-1}_{\nu_1,\nu_2} 
 \:\:\:.
\nonumber
 \end{eqnarray}
 \noindent
 Once the $\theta_{{\bf k},\nu}$ modes are pertinently integrated over, the free energy of our system must 
 be quadratic in the ${\cal Q}_{x,y}$. To evidence this, we trade $\hat{H}_{{\rm Vil},2}^{\rm mod} [\vec{\cal Q} ]$
 [that is, $H_{{\rm Vil},2}^{\rm mod} [\vec{\cal Q} ]$
 with all the $J^{x(y)}(\nu)$ substituted with $\hat{J}_{x(y)}(\nu,T)$] for the effective 
 Villain Hamiltonian $H_{{\rm Eff},{\rm mod}}^{\rm Vil} [\vec{\cal Q}]$, defined  (apart for an unessential constant) 
 via a systematic 
 integration over the $\theta_{{\bf k},\nu}$ modes, with $\nu \neq 0$, according to 
 
 \beq
 e^{-\frac{H_{{\rm Eff},{\rm mod}}^{\rm Vil} [\vec{\cal Q}]}{T}} 
 = \int \prod_{{\bf k}} \prod_{\nu=1}^{N-1} d \theta_{{\bf k},\nu} e^{-\frac{
 \hat{H}_{{\rm Vil},2}^{\rm mod} [\vec{\cal Q} ]}{T}}
 \:\: .
 \label{neweq.1}
 \eneq
 \noindent
 As a result, we obtain Eq.(\ref{mxz.1}) of the main text, with 
 
\begin{eqnarray}
&& {\cal J}^x(T) = \hat{J}^x ( 0 ,T) \equiv {\cal J}^x_0-T{\cal J}^x_1 \; ,   \nonumber \\
&& {\cal J}^y(T) =\hat{J}^y( 0,T)-\sum_{\nu_1,\nu_2=1}^{N-1} \hat{J}^y (\nu_1,T ) (1-e^{-\frac{2 \pi i \nu_1}{N}}) \nonumber \\
&&  \times [\tilde{\cal D} ({\bf 0}) ]_{\nu_1,\nu_2}^{-1} (1-e^{\frac{2 \pi i \nu_2}{N}}) \hat{J}^y (-\nu_2,T) \equiv {\cal J}^y_0-T{\cal J}^y_1 
\:\:\: . 
\label{mxy.14}
\end{eqnarray}
\noindent
Setting $\delta_{x(y)}={\cal J}_0^{x(y)}/{\cal J}_1^{x(y)}$ yields Eq.(\ref{mxy.15}) of the main text. 

The kernel $\Delta ({\bf k})$ in Eq.(\ref{mxz.1}) is defined as 

\beq
\Delta ({\bf k}) =\hat{\cal D}_{0,0}({\bf k}) -\sum_{\nu,\nu'=1}^{N-1}\hat{\cal D}_{0,\nu}({\bf k}) 
[ \tilde{\cal D}^{-1} ({\bf k}) ]_{\nu,\nu'} \hat{\cal D}_{\nu',0} ({\bf k}) 
\:\: ,
\label{mxz.2}
\eneq
\noindent
with   $\tilde{\cal D} ({\bf k})$ in Eq. (\ref{mxy.14}) being an $(N-1)\times (N -1)$ matrix, obtained
from $\hat{\cal D} ({\bf k})$ by dropping the first row and the first column, with 
$\hat{\cal D}_{\nu,\nu'}({\bf k})$   equal to  ${\cal D}_{\nu,\nu'} ({\bf k})$ in Eq. (\ref{mxy.11}), and 
with $J^{x(y)}(\nu)$ substituted with $\hat{J}^{x(y)}(\nu,T)$.  Expanding $\Delta ({\bf k})$ around 
${\bf k}={\bf 0}$ up to second order in ${\bf k}$, we obtain Eq.(\ref{mxz.3}) of the main text.

   \bibliography{bib_anisotropic}
\end{document}